%% file: llncs-main.tex
\documentclass[letterpaper]{llncs}
\usepackage[numbers,compress]{natbib}
\usepackage{graphicx}
\usepackage{booktabs}
\usepackage{amsmath}
\usepackage{amsfonts}
\usepackage{amssymb}

\usepackage[ruled,linesnumbered,noend]{algorithm2e}
\usepackage{caption,subcaption}

\SetCommentSty{mycommfont}

\SetKwRepeat{Do}{do}{while}

\usepackage{multicol}
\usepackage{latexsym}
\usepackage{multirow}
\usepackage{threeparttable}
\usepackage{pifont}
\usepackage{balance}
\usepackage{float}
\usepackage{url}
\usepackage{hyperref}
\hypersetup{colorlinks=true, linkcolor=blue, citecolor=violet, filecolor=magenta}
\usepackage{tcolorbox}
\usepackage{galois}
\usepackage[utf8]{inputenc}
\usepackage{cleveref}
\crefname{section}{§}{§§}
\Crefname{section}{§}{§§}
\usepackage{stmaryrd}
\usepackage{tikz}
\usetikzlibrary{arrows, automata}
\usepackage{pgfplots}
\pgfplotsset{compat=1.3}

\newcommand{\mytodoblue}[1]{\textcolor{blue}{\ding{46}~{\sf}~#1}}
\newcommand{\mytodored}[1]{\textcolor{red}{\ding{46}~{\sf}~#1}}

\newif\ifshowcomments
\showcommentstrue

\ifshowcomments
\newcommand{\yao}[1]{\mytodoblue{[yao: #1]}}
\newcommand{\zheng}[1]{\mytodored{[zheng: #1]}}
\else
\newcommand{\yao}[1]{}
\newcommand{\zheng}[1]{}
\fi

\newcommand{\mybox}[1]{
    \begin{tcolorbox}[
	boxsep=-0.5pt,
	standard jigsaw,
	boxrule=0.6pt,
	opacityback=0,
	sharp corners]
	#1
    \end{tcolorbox}
}

\usepackage{xcolor}

\newcounter{finding}
\newcommand{\finding}{%
  \stepcounter{finding}%
  \textbf{Finding} \textbf{\arabic{finding}}:%
}

\begin{document}

\title{Analyzing the Analyzers: Model Counting Meets Abstract Interpretation}

\author{Junda Zheng  \and Peisen Yao} \institute{The State Key Laboratory of Blockchain and Data Security, Zhejiang University \\ \email{\{zhengjd04,pyaoaa\}@zju.edu.cn}}

\maketitle

\begin{abstract}	
Abstract interpretation offers a principled foundation for static analysis by approximating concrete program semantics via abstract domains. However, quantitatively comparing the precision of different domains remains a longstanding challenge.
We present MCAI (Model Counting meets Abstract Interpretation), a new methodology that employs model counting to measure the precision of abstract domains. Unlike prior approaches that assess precision relative to specific analysis queries, MCAI encodes both concrete semantics and abstract values as logical formulas, enabling a client-independent, quantitative metric of imprecision that captures the inherent semantic loss in the abstractions.
We apply MCAI to four abstract domains and evaluate the precision of their best abstract transformers via symbolic abstraction. Our results yield several insights: the Interval domain, despite its simplicity, often achieves precision comparable to that of Octagon; many octagonal constraints are semantically redundant; and the bit-level KnownBit domains consistently outperform the word-level domains.  MCAI offers both theoretical insights into the precision of abstract domains and practical guidance for selecting suitable abstractions.
\end{abstract}
	
\keywords{Abstract interpretation, model counting, evaluation}
\input{1.intro}
\input{2.background}

\input{3.overview}

\input{4.method} 
\input{5.eval}

\input{6.discuss}

\input{7.related}
\input{8.conclu}

\footnotesize
\bibliographystyle{unsrtnat}
\bibliography{sigproc}
\end{document}

%% file: 1.intro.tex
\section{Introduction}
\label{sec:intro}

Abstract interpretation~\cite{cousot1976static, cousot1978automatic} is a foundational framework that enables sound reasoning over potentially infinite sets of concrete program states by mapping them into abstract values in an abstract domain. 
An abstract domain encodes approximations of concrete states, enabling tractable static analyses.
Given a Galois connection between a concrete domain $C$ and an abstract domain $A$, the best abstract transformer $f^\#: A \to A$ is $f^\# = \alpha \circ f \circ \gamma$, where $f$ is the concrete transformer, and $\alpha$ and $\gamma$ are the abstraction and concretization functions, respectively. 
This definition establishes the theoretical upper bound on the precision achievable within a given abstract domain.

Despite significant advancements in abstract interpretation, a critical challenge remains: How can we objectively evaluate and compare the precision of abstract transformers? Consider a program fragment with constraints $x \geq 0 \land y \geq 0 \land x + y \leq 10$, where $x$ is an integer variable.
The interval domain approximates this as $x \in [0, 10] \land y \in [0, 10]$. 
The octagon domain can express $x \geq 0 \land y \geq 0 \land x + y \leq 10$ precisely.
While the octagon domain is more precise in this case, we lack systematic methods to quantify how much more accurate it is, especially when dealing with complex bit-precise semantics.

Quantitative evaluation of abstract domains is essential for tracking scientific progress and guiding the development of practical tools. Benchmarking enables comparative analysis across domains and supports informed engineering decisions. Existing benchmark suites have been widely adopted for evaluating static analysis tools, such as DaCapo~\cite{blackburn2006dacapo} and OWASP~\cite{owasp}. Prior work has also explored domain-specific metrics for assessing the precision of abstract transformers~\cite {logozzo2009towards, di2008relational, sotin2010quantifying}. However, these efforts suffer from two key limitations:
\begin{itemize}
    \item \emph{Client-Specific Evaluation Metrics}: Most evaluation techniques are tied to specific downstream tasks, such as program verification (e.g., finding numerical invariants), program understanding (e.g., slicing), and compiler optimizations (e.g., devirtualization). This coupling makes it difficult to draw general conclusions about domain precision that hold across different contexts.
    \item \emph{Limited Quantification w.r.t Concrete Semantics}: Existing methods rarely measure the extent to which abstract domains over-approximate concrete semantics, e.g., providing exact measurements of information loss during abstraction. When concrete semantics are considered, evaluations typically rely on incomplete under-approximations based on a finite set of dynamic executions, which do not capture the entire behavior space.
\end{itemize}

In response to these limitations, we introduce MCAI (Model Counting meets Abstract Interpretation), a methodology for quantitatively evaluating the precision of abstract domains. MCAI encodes both concrete semantics and abstract values as logical formulas and leverages model counting techniques to measure the extent to which abstract representations over-approximate the concrete behavior space (\cref{subsec:precison_def}). Additionally, we introduce a formal metric for quantifying semantic differences among abstract domains (\cref{subsec:comparison}).

To demonstrate the effectiveness of MCAI, we apply it to numerical abstract domains, where concrete program semantics are encoded as quantifier-free bit-vector formulas. Specifically, we analyze the precision of the best abstract transformers via symbolic abstraction \cite{reps2004symbolic}, which provides the most precise over-approximation of concrete operations within a given domain. 
Our experimental evaluation is based on SMT queries from two well-known program analyzers—CBMC and Kint—and focuses on four numerical abstract domains: Interval \cite{cousot1976static}, Zone \cite{Mine:Zone}, Octagon \cite{mine2006octagon}, and KnownBit. Our experimental evaluation reveals several interesting—and sometimes counterintuitive—findings. For example:
\begin{itemize}
    \item  A common belief is that more expressive domains, such as Octagons, inherently offer superior precision. We find that the Interval domain often achieves precision comparable to that of the Octagon domain, with average precision scores of 76.1\% versus 77.2\%, respectively.
    \item Many constraints within the Octagon domain are redundant, particularly the ``plus'' constraints, which contribute negligibly to precision improvement while significantly increasing computational overhead.
    \item The KnownBit domain consistently outperforms word-level domains, achieving a mean precision of 85.7\% compared to 76-77\% for numerical domains, suggesting that bit-level reasoning captures structural properties that numerical abstractions miss.
    \item Model counting over bit-vector formulas proves tractable for practical analysis, with median analysis times ranging from 0.48 seconds for Interval to 4.65 seconds for Octagon on our benchmark suite of 2,006 formulas.
\end{itemize}

To summarize, this paper makes the following main contributions:

\begin{itemize}
    \item We develop a model-counting-based methodology to quantitatively evaluate the precision of abstract domains and their semantic differences relative to concrete semantics.
    \item We systematically compare four abstract domains, revealing new insights into their relative precision and computational trade-offs. 
    \item We make our tool publicly available at \url{https://tinyurl.com/yxvn6a5n}. 
\end{itemize}

%% file: 2.background.tex
\section{Preliminaries}
\label{sec:overview}

\noindent \textbf{Abstract Interpretation}.
Our work is based on the Galois connection framework within the context of abstract interpretation. 
Given two complete lattices $(C, \leq_C)$ and $(A, \leq_A)$, a pair of functions—an abstraction function $\alpha : C \to A$ and a concretization function $\gamma : A \to C$—forms a \textit{Galois connection} if, for any element $c \in C$ and $a \in A$, $\alpha(c) \leq_A a \Leftrightarrow c \leq_C \gamma(a)$.
This relationship between the concrete domain $C$ and the abstract domain $A$ ensures that reasoning in the abstract domain safely over-approximates behaviors in the concrete domain.

\smallskip
\noindent \textbf{Best Abstract Transformer}. Given a concrete transformer $f : C \to C$, the \emph{best abstract transformer} $f^\alpha : A \to A$ that over-approximates $f$ is defined as:
$
    f^\alpha = \alpha \circ f \circ \gamma : A \to A.
$
This is the most precise sound abstraction of $f$ in the abstract domain $A$ since, for any other sound abstraction $f^\#$, it holds that $f^\alpha(a) \leq_A f^\#(a)$ for all $a \in A$.
\footnote{A best abstract transformer may not  exist~\cite{cousot1978automatic,cousot1995formal,cousot2011logical}. In such cases, only one direction of the Galois connection may be used (e.g., the concretization function only). This work focuses on abstract domains where the best abstract transformers exist.}
Despite its theoretical significance, this definition is \textit{non-constructive}, meaning it does not necessarily provide an algorithm to compute the most precise transfer function $f^\alpha$. Furthermore, the composition of the best abstractions of two functions $f$ and $g$ does not always yield the best abstraction of their composition $f \circ g$.

\smallskip
\noindent \textbf{Computing the Best Abstraction}. 
A practical method to compute the best abstraction is \textit{symbolic abstraction}~\cite{reps2004symbolic}, which finds the strongest consequence of a formula $\varphi \in \mathcal{L}$ that is expressible within an abstract domain $A$. Specifically, given a formula $\varphi$ representing the concrete semantics and an abstract domain $A$, symbolic abstraction computes the best approximation of $\varphi$ in $A$.
Depending on the clients, the formula $\varphi \in \mathcal{L}$ may encode different language constructs,  such as the concrete transformer for an instruction, a basic block, or a loop-free program fragment.
Symbolic abstraction is beneficial in two key ways: it facilitates the automatic synthesis of optimal transformers and mitigates precision loss when composing multiple transformers.

   \begin{example}
    Consider the interval domain over integers. The logic fragment corresponding to this domain, $\mathcal{L}_{Interval}$,
    comprises conjunctions of one-variable inequalities over the program variables. The full logic might be the theory of quantifier-free non-linear integers (QF\_NIA) or linear integers (QF\_LIA).
\end{example} 

%% file: 3.overview.tex
\section{Problem Statement}
\label{sec:problemstatement}

Abstract domains provide sound over-approximations of program behavior, but this soundness comes at the cost of precision. Quantifying the precision of abstract transformers is therefore essential for evaluating and improving static analyses grounded in abstract interpretation. A key challenge is measuring the loss of precision and comparing domains in a principled, domain-agnostic manner~\cite{sotin2010quantifying, logozzo2009towards}.
Existing precision metrics have two key limitations. First, they are often client-specific, entangling domain precision with analysis behavior. Second, they typically assess abstract elements in isolation or rely on dynamic traces, which offer limited semantic coverage.

This paper proposes \textit{Model Counting Meets Abstract Interpretation} (MCAI), a methodology that enables such an evaluation by combining logical encoding with model counting.
MCAI has two primary objectives:
\begin{itemize}
    \item \textbf{Client-Independent Quantification}: a  general-purpose metric for abstraction loss that does not rely on specific analysis goals or client heuristics;
    \item \textbf{Concrete-Semantics Alignment}: evaluating how closely an abstract transformer approximates the behavior of its corresponding concrete transformer.
\end{itemize}

This approach offers several key benefits. 
Evaluations grounded directly in program semantics, rather than being influenced by the success or failure of downstream verification tasks;
Systematic assessment of enhancements to abstract domains, such as finite disjunctions, refined join or widening operations, or increased expressiveness, by quantifying their impact on precision.
For instance, instead of stating that an abstract interpreter verified 85\% of assertions in a benchmark suite, we can report that a particular domain exhibits a 13\% loss of precision with respect to the concrete semantics. This provides an objective and transferable measure of domain quality.

To illustrate and validate this approach, we focus on bit-precise static analysis. Program semantics are represented as quantifier-free formulas over bit-vectors (QF\_BV), which accurately model low-level integer operations. Specifically,  we address the following problem:
\mybox{
Given a formula $\varphi$ and a set of numeric domains $\{ A_1,  A_2, \ldots, A_n \}$ over fixed-size integers, (i) provide an automatic method to measure the precision of abstractions of $\varphi$ in each domain $A_i$; (ii) compare the semantic differences between these abstract domains.}

%% file: 4.method.tex
\section{Model Counting Meets Abstract Interpretation}
\label{sec:mcai}
This section introduces the MCAI framework, which leverages symbolic characterizations of both concrete and abstract semantics, along with model counting, to assess and quantify the precision of abstract domains. 
We first define the semantics of programs and abstract elements, then introduce model counting as a mechanism for precision measurement, and finally show how these notions provide a new lens on completeness in abstract interpretation.

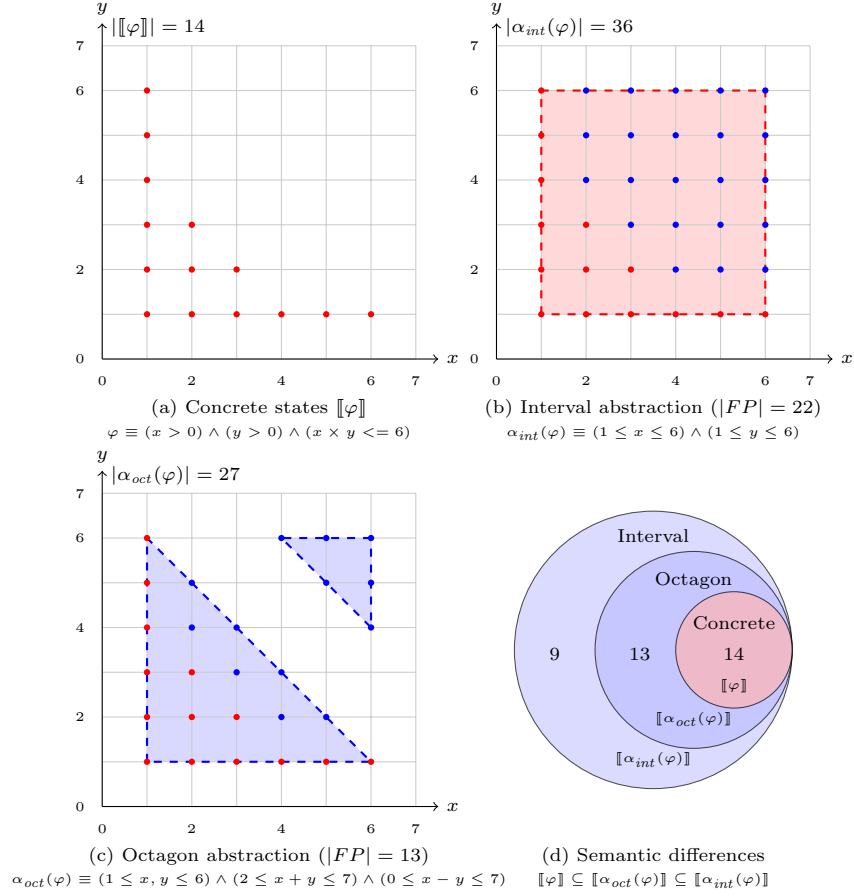
\begin{figure}[htbp]
\centering
\input{images/idea.tikz}
\caption{Illustration of an example of a concrete semantics formula and its best abstraction in the interval and octagon domain. With 3-bit unsigned bit-vector arithmetic, an unexpected false positive area in the octagon domain is caused by addition overflow. The underflow of subtraction also influences $x-y$.}
\label{fig:sound_comparison}
\end{figure}

\subsection{Formalizing and Measuring Abstraction Precision}
\label{subsec:precison_def}
To measure abstraction precision, we begin by representing both concrete program semantics and abstract elements as logical formulas. This standard representation allows us to analyze their relationships using model-theoretic tools.

\begin{definition}
Let $\varphi$ be a formula representing the concrete semantics of a program. The set of concrete states satisfying $\varphi$ is denoted by $\llbracket\varphi\rrbracket = \{M \mid M \models \varphi\}$, where $M$ is a truth assignment over the variables in $\varphi$.
\end{definition}

\begin{example}
Consider a program with two integer variables $x$ and $y$. The concrete semantics $\varphi(x,y) \equiv x \geq 0 \land y \geq 0 \land x + y \leq 5$ represents all valid states where both variables are non-negative, and their sum does not exceed five.
\end{example}

\smallskip
\noindent \textbf{Symbolic Concretization}.  
While concrete semantics define exact behaviors, abstract domains approximate these behavior sets. To facilitate comparison, we introduce symbolic concretization—an approach that maps abstract elements back into logical formulas over program variables.

\begin{definition}
\label{def:symbolic-concretization}
Let $\mathcal{L}_{A}$ be a fragment of logic $\mathcal{L}$. A symbolic concretization function $\gamma$ maps each abstract element $a \in A$ to a formula $\varphi_a \in \mathcal{L}_{A}$ such that $\llbracket\varphi_a\rrbracket = \gamma(a)$. Since $\mathcal{L}_{A} \subseteq \mathcal{L}$, the formula $\varphi_a$ may also be interpreted in full logic.
\end{definition}

\begin{example}
Intuitively, symbolic concretization maps an abstract element to a logical formula that characterizes its set of concrete states.
For the interval domain, an abstract element $a$ representing $x \in [0, 10]$ has symbolic concretization $\varphi_a \equiv 0 \leq x \leq 10$. This formula precisely captures the set of concrete states represented by the interval.
\end{example}

This logical representation enables us to reason about abstract elements using satisfiability and model-counting tools, allowing a quantitative comparison between abstract and concrete state spaces.

\smallskip
\noindent \textbf{Counting Measure}.  
We quantify how much abstract elements over-approximate the concrete semantics when expressed as formulas by counting the models of each formula.

\begin{definition} 
	Given an abstract element $S$, the counting measure is defined as:
	\begin{equation}
		vol(S) = \left\{
		\begin{aligned}
			| S | & , & \text{if} \ S \ \text{is finite}, \\
			+\infty  & , & \text{otherwise}.
		\end{aligned}
		\right.
	\end{equation}
\end{definition}

This measure is particularly applicable to domains with bounded types, such as Booleans and fixed-width bit-vectors.

\begin{example}
Consider the formula $\varphi(x,y) \equiv x \geq 0 \land y \geq 0 \land x \times y \leq 10$. An interval abstraction like $x \in [0, 10] \land y \in [0, 10]$ over-approximates this region.
However, the interval includes infeasible states such as $(x = 10, y = 10)$, which violate the original constraint.
\end{example}

\noindent \textbf{Measuring the Precision}.
\label{subsec:precision_quant}
To make imprecision explicit, we define false positives as states that satisfy the abstraction but violate the original semantics.

\begin{definition}
Let $\varphi$ be a concrete semantics formula, and let $\alpha_A(\varphi)$ be its over-approximation in the abstract domain $A$. A model $M$ is a false positive if $M \models \alpha_A(\varphi)$ and $M \not\models \varphi$. That is, $M$ satisfies 
$$\alpha_A(\varphi) \land \neg \varphi$$
\end{definition}

\begin{definition}  [Number of False Positives]
    The number of false positives for an abstraction $\alpha_A(\varphi)$ in domain $A$:
	\begin{align*}
		FP(A) = |\{M \ | \ M \models  \alpha_A(\varphi) \land \neg \varphi\}|.
	\end{align*}
\end{definition}

\begin{example}
Consider the formula $\varphi(x,y) \equiv x > 0 \land y > 0 \land x \times y \leq 6$. The interval abstraction $\alpha_{int}(\varphi) \equiv 1 \leq x \leq 6 \land 1 \leq y \leq 6$ includes points such as $(6,6)$ that do not satisfy the original formula $\varphi$. Such points are counted as false positives. As shown in Figure~\ref{fig:sound_comparison}, this interval abstraction introduces 22 false positive states.
\end{example}

Intuitively, this discrepancy between the concrete and abstract state counts represents the degree of imprecision.

\subsection{Comparative Precision Between Abstract Domains}
\label{subsec:comparison}
After quantifying the imprecision of individual abstractions, we now explore how different abstract domains compare in their over-approximation behavior.

\begin{definition} [Abstract Domain-Specific False Positives]
	\label{def:diff}
Let $A$ and $B$ represent two abstract domains. For a given formula $\varphi$, the false positives can be categorized as:
	\begin{itemize}
		\item \emph{Common false positives}: States that are false positives in both domains:
          \begin{equation}
              \alpha_A(\varphi) \land \neg \varphi \land \alpha_B(\varphi)
          \end{equation}
		\item \emph{$A$-specific false positives}:  States that are false positives in $A$ but not in $B$: 
          \begin{equation}
              \alpha_A(\varphi) \land \neg \varphi \land \neg \alpha_B(\varphi)
          \end{equation}
        
		\item \emph{$B$-specific false positives}: States that are false positives in $B$ but not in $A$: 
        \begin{equation}
          \neg \alpha_A(\varphi) \land \neg \varphi \land \alpha_B(\varphi)  
        \end{equation}
	\end{itemize}
\end{definition}

\begin{example}
Using the earlier formula $\varphi$, suppose $\alpha_{interval}(\varphi)$ and $\alpha_{octagon}(\varphi)$ produce different abstractions. We can quantitatively assess their relative precision by computing false positives using the logical formulas defined above, as illustrated in Figure~\ref{fig:sound_comparison}. The analysis reveals 13 common false positives shared between both domains and 9 false positives specific to the interval domain. Notably, no false positives are unique to the octagon domain. 
\end{example}

\section{Implementations and Applications}
\label{sec:impl}

\noindent \textbf{Implementations}.
We have implemented MCAI as a modular analysis framework comprising nearly 5,000 lines of Python code. MCAI operates by encoding the concrete semantics and abstracting them as logical formulas.

Figure~\ref{fig:architecture} illustrates MCAI's architecture, consisting of two main components:

\begin{figure}[t]
    \centering
    \includegraphics[width=0.85\linewidth]{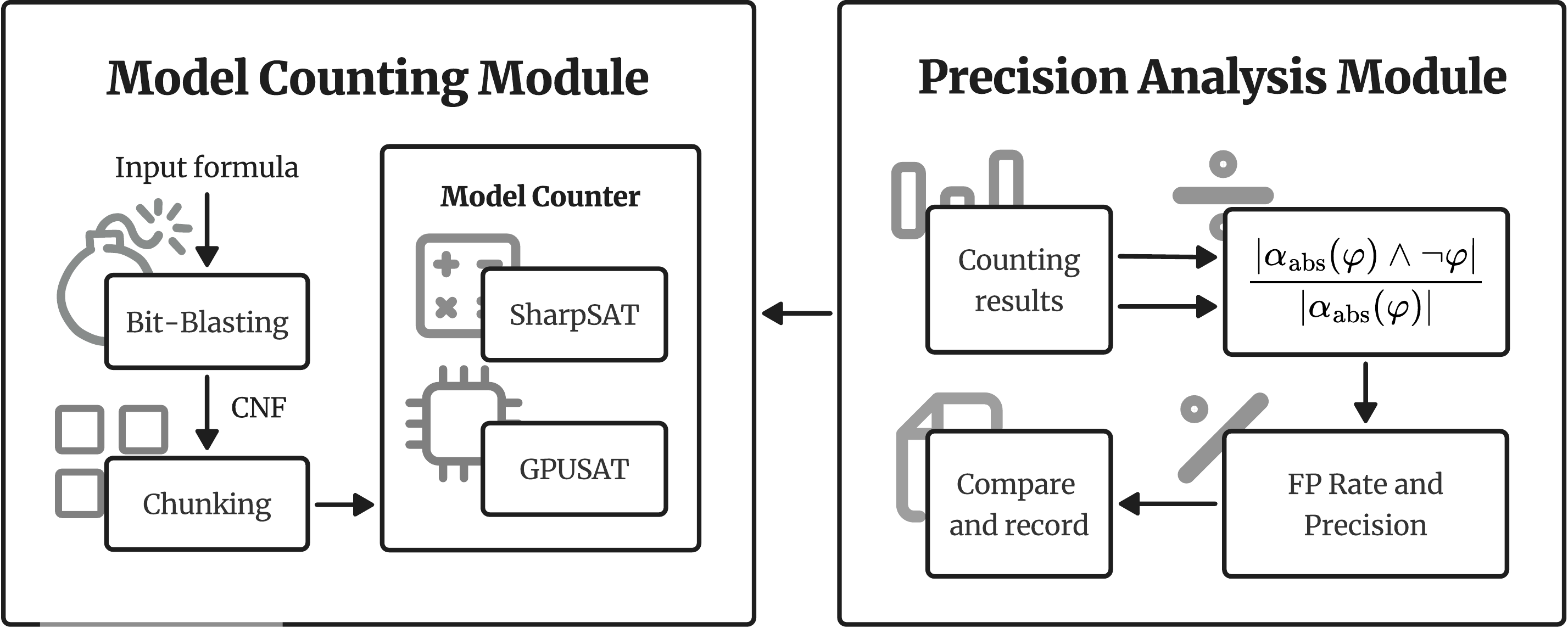}
    \caption{Illustration of the architecture and overall workflow of MCAI.}
    \label{fig:architecture}
\end{figure}

\begin{itemize}
    \item \emph{Model Counting}: This module performs model counting over logical formulas. We bit-blast the bit-vector formulas to propositional logic and apply off-the-shelf model counters. To mitigate the distortion introduced by auxiliary variables during bit-blasting and CNF conversion, we employ projected model counting, restricting the count to the priority variables.
    \item \emph{Precision Analysis}. This component quantifies the precision of abstract domains by comparing the model counts of their abstractions against the concrete semantics. The resulting metrics enable a systematic evaluation of the expressiveness of abstract domains.

\end{itemize}

\noindent \textbf{Applications}.
MCAI supports a range of use cases that facilitate deeper understanding and empirical evaluation of abstract domains. In our evaluation (\cref{sec:eval}), we showcase several practical applications:
\begin{itemize}
    \item \emph{Precision of Best Abstraction} (\cref{subsec:rq1}). We compute the most accurate over-approximations within a given abstract domain and use model counting to quantify deviation from concrete semantics. This analysis reveals the theoretical limits of precision achievable by different abstract domains.
    \item \emph{Comparative Analysis of Domains} (\cref{subsec:rq2}). 
     We perform comparisons of domain effectiveness across multiple abstract domains, providing quantitative rankings of their relative precision. 
     These results may assist in the evidence-based selection of abstract domains for specific analysis scenarios.
   \item \emph{Constraint Redundancy Analysis} (\cref{subsec:rq3}). We measure individual constraint contributions to the precision, such as the plus constraints in the Octagon domain. This analysis helps identify which constraints are important for maintaining precision and which may be omitted to improve performance.

\end{itemize}  

%% file: images/idea.tikz
\begin{tikzpicture}[scale=0.592, every node/.style={font=\scriptsize}]

\begin{scope}[xshift=-4.4cm]
    \draw[gray!40] (0,0) grid (7,7);
    \draw[->] (0,0) -- (7.5,0) node[right]{$x$};
    \draw[->] (0,0) -- (0,7.5) node[above]{$y$};
    
    \foreach \i in {0,2,4,6,7} {
        \node[below] at (\i,-0.2) {\tiny \i};
        \node[left] at (-0.2,\i) {\tiny \i};
    }
    
    \foreach \x in {1,2,...,7} {
        \foreach \y in {1,2,...,7} {
            \pgfmathparse{int(\x*\y<=6 ? 1 : 0)}
            \ifnum\pgfmathresult=1
                \fill[red] (\x,\y) circle (2pt);
            \fi
        }
    }
    
    \node[below] at (3.5,-0.7) {(a) Concrete states $\llbracket\varphi\rrbracket$};
    \node[above right] at (0,7) {$|\llbracket\varphi\rrbracket| = 14$};
    \node[below] at (3.5,-1.3) {\tiny $\varphi \equiv (x > 0) \wedge (y > 0) \wedge (x\times y <= 6)$};
\end{scope}

\begin{scope}[xshift=4.4cm]
    
    \foreach \i in {0,2,4,6,7} {
        \node[below] at (\i,-0.2) {\tiny \i};
        \node[left] at (-0.2,\i) {\tiny \i};
    }
    
    \fill[red!15] (1,1) rectangle (6,6);

    \draw[gray!40] (0,0) grid (7,7);
    \draw[->] (0,0) -- (7.5,0) node[right]{$x$};
    \draw[->] (0,0) -- (0,7.5) node[above]{$y$};

    \draw[dashed,red,thick] (1,1) rectangle (6,6);
    
    \foreach \x in {1,2,...,7} {
        \foreach \y in {1,2,...,7} {
            \pgfmathparse{int(\x*\y<=6 ? 1 : 0)}
            \ifnum\pgfmathresult=1
                \fill[red] (\x,\y) circle (2pt);
            \fi
        }
    }
    
    \foreach \x in {1,2,...,6} {
        \foreach \y in {1,2,...,6} {
            \pgfmathparse{int(\x*\y>6 ? 1 : 0)}
            \ifnum\pgfmathresult=1
                \fill[blue] (\x,\y) circle (2pt);
            \fi
        }
    }
    
    \node[below] at (3.5,-0.7) {(b) Interval abstraction ($| FP | =22$)};
    \node[above right] at (0,7) {$|\alpha_{\mathit{int}}(\varphi)| = 36$};
    \node[below] at (3.5,-1.3) {\tiny $\alpha_{\mathit{int}}(\varphi) \equiv (1 \leq x \leq 6) \wedge (1 \leq y \leq 6)$};
\end{scope}

\begin{scope}[xshift=-4.4cm, yshift=-10cm]
    
    \foreach \i in {0,2,4,6,7} {
        \node[below] at (\i,-0.2) {\tiny \i};
        \node[left] at (-0.2,\i) {\tiny \i};
    }
    
    \fill[blue!15] (1,1) -- (1,6) -- (6,1) -- cycle;
    \fill[blue!15] (6,6) -- (4,6) -- (6,4) -- cycle;

    \draw[gray!40] (0,0) grid (7,7);
    \draw[->] (0,0) -- (7.5,0) node[right]{$x$};
    \draw[->] (0,0) -- (0,7.5) node[above]{$y$};

    \draw[dashed,blue,thick] (1,1) -- (1,6) -- (6,1) -- cycle;
    \draw[dashed,blue,thick] (6,6) -- (4,6) -- (6,4) -- cycle;
    
    \foreach \x in {1,2,...,7} {
        \foreach \y in {1,2,...,7} {
            \pgfmathparse{int(\x*\y<=6 ? 1 : 0)}
            \ifnum\pgfmathresult=1
                \fill[red] (\x,\y) circle (2pt);
            \fi
        }
    }
    
    \foreach \x in {1,...,7} {
        \foreach \y in {1,...,7} {
            \pgfmathparse{int((\x*\y<=6) ? 0 : ((\x<=6 && \y<=6 && (\x+\y>=2 && \x+\y<=7 || \x+\y>=10 && \x+\y<=15)) ? 1 : 0))}
            \ifnum\pgfmathresult=1
                \fill[blue] (\x,\y) circle (2pt);
            \fi
        }
    }
    
    \node[below] at (3.5,-0.7) {(c) Octagon abstraction ($| FP |=13$)};
    \node[above right] at (0,7) {$|\alpha_{\mathit{oct}}(\varphi)| = 27$};
    \node[below] at (3.5,-1.3) {\tiny $\alpha_{\mathit{oct}}(\varphi) \equiv (1 \leq x,y \leq 6) \wedge (2 \leq x + y \leq 7) \wedge (0 \leq x - y \leq 7)$};
\end{scope}

\begin{scope}[yshift=-10cm, xshift=4.4cm]
%
%
%
%
%
%
    
    \draw [fill=blue!20,opacity=0.7] (3.5,3.5) circle (3.1cm);
    \draw [fill=blue!25,opacity=0.7] (4.4,3.5) circle (2.2cm);
    \draw [fill=red!30,opacity=0.7] (5.3,3.5) circle (1.3cm);
    \node at (1.3,3.4) {\scriptsize $9$};
    \node at (3.2,3.4) {\scriptsize $13$};
    \node at (5.3,3.4) {\scriptsize $14$};
    \node at (3.5,6.05) {\scriptsize Interval};
    \node at (4.4,5.05) {\scriptsize Octagon};
    \node at (5.3,4.1) {\scriptsize Concrete};
    \node at (3.5,1.05) {\tiny $\llbracket\alpha_{\mathit{int}}(\varphi)\rrbracket$};
    \node at (4.4,1.95) {\tiny $\llbracket\alpha_{\mathit{oct}}(\varphi)\rrbracket$};
    \node at (5.3,2.7) {\tiny $\llbracket\varphi\rrbracket$};
    \node [below] at (3.5,-0.7) {(d) Semantic differences};
    \node [below] at (3.5,-1.3) {\tiny $\llbracket\varphi\rrbracket \subseteq \llbracket\alpha_{\mathit{oct}}(\varphi)\rrbracket \subseteq \llbracket\alpha_{\mathit{int}}(\varphi)\rrbracket$};

\end{scope}

\end{tikzpicture}

%% file: 5.eval.tex
\section{Evaluation}
\label{sec:eval}

This section presents our evaluation 
by studying the following research questions:
\begin{itemize}
    \item \textbf{RQ1}: What is the precision of best abstractions with respect to concrete semantics across different domains (\cref{subsec:rq1})?
    \item \textbf{RQ2}: How do different abstract domains compare quantitatively in terms of precision and false positive rates (\cref{subsec:rq2})?
    \item \textbf{RQ3}: Which categories of constraints in relational domains contribute most to precision improvements (\cref{subsec:rq3})?
    \item \textbf{RQ4}: What is the computational overhead of MCAI when computing best abstractions and counting models (\cref{subsec:rq5})?
\end{itemize}

\noindent \textbf{Benchmark Dataset}.
We evaluate MCAI using a benchmark of 2,006 bit-vector formulas collected from two widely-used program analyzers:
\begin{itemize}
    \item CBMC~\cite{cbmc}:  A bounded model checker for ANSI C programs that is used extensively in academia and industry; 
    \item KINT~\cite{kint}: A tool that uses path-sensitive static analysis to detect runtime memory errors in C/C++ programs. 
\end{itemize}

\begin{figure}[t]
    \centering
    \includegraphics[width=0.85\linewidth]{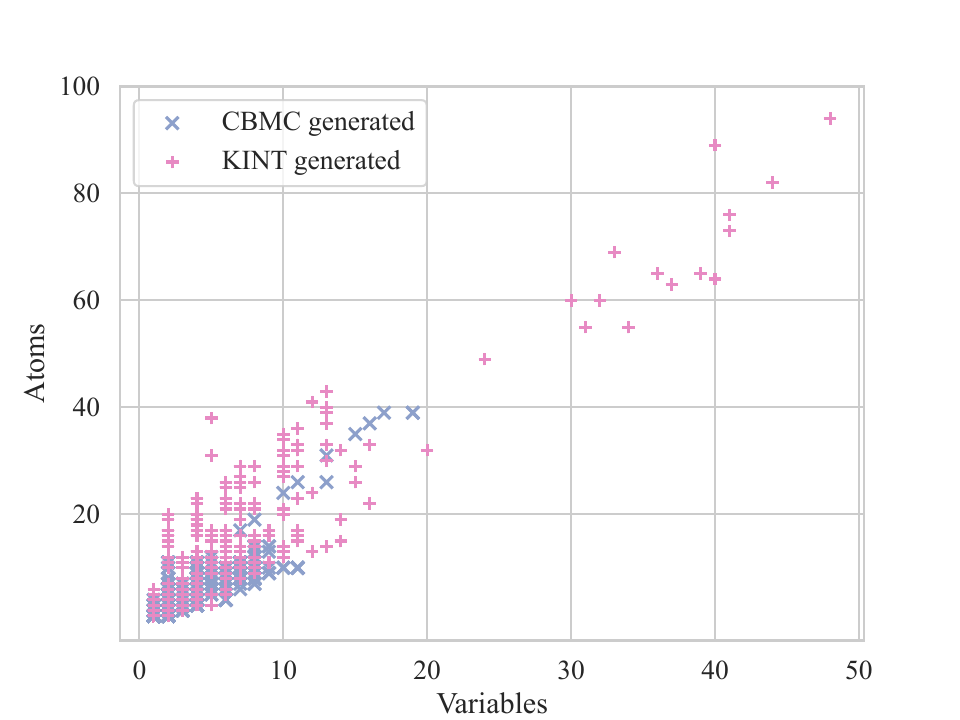}
    \caption{Distribution of variables and atomic predicates across the benchmark formulas. Most formulas contain 1-20 variables and 1-40 atomic predicates.}
    \label{fig:varatom}
\end{figure}

Our benchmarks were drawn from the Reach Safety Track of SV-COMP. To keep analysis times tractable, we excluded formulas that cause timeouts during model counting. In total, we generated 3,100 formulas from SV-COMP, filtered out 1094 instances, and retained 2006 for evaluation. Although this filtering reduces the total benchmark pool, it preserves a wide variety of formula structures. 
Tables~\ref{eval:cbmc-formula} and~\ref{eval:kint-formula} summarize the dataset composition.
Figure~\ref{fig:varatom} shows the distribution of variable and predicate counts.  Most benchmark formulas fall within a moderate complexity range (1–20 variables and 1–40 predicates), yet they still exhibit wide variation in solution space sizes—from a single satisfying model to as many as $2^{479}$ models (with a median of $2^{112}$). 

\begin{minipage}[t]{0.46\linewidth}
    \centering
    \captionof{table}{The number of formulas collected from CBMC in each subdirectory of SV-COMP}
    \label{eval:cbmc-formula}
    \begin{tabular*}{\linewidth}{@{\extracolsep{\fill}} l c }
        \toprule
	\textbf{Subdirectory} & \textbf{Number} \\ \midrule
	bitvector-regression 		& 80 \\
	list-ext2-properties 		& 156 \\
        termination-bwb 		& 296 \\
        termination-crafted     	& 98 \\
        termination-numeric     	& 101 \\
        termination-recursive-malloc    & 162 \\
        \bottomrule
    \end{tabular*}
\end{minipage}
\hfill
\begin{minipage}[t]{0.46\linewidth}
    \centering
    \captionof{table}{The number of formulas collected from KINT in each subdirectory of SV-COMP}
    \label{eval:kint-formula}
    \begin{tabular*}{\linewidth}{@{\extracolsep{\fill}} l c }
        \toprule
	\textbf{Subdirectory} & \textbf{Number} \\ \midrule
	bitvector-regression 		& 74 \\
	ldv-validator-v0.6		& 480 \\
	list-ext-properties		& 99 \\
	list-ext2-properties 		& 74 \\
	product-lines			& 28 \\
        termination-bwb 		& 68 \\
        termination-recursive-malloc    & 208 \\
	validation-crafted		& 82 \\
        \bottomrule
    \end{tabular*}
\end{minipage}

\medskip
\noindent \textbf{Abstract Domains}.
We instantiate MCAI using the following word- and bit-level abstract domains:
\begin{itemize}
    \item Interval~\cite{cousot1976static}: A non-relational domain that handles single-variable inequalities of the form $a \leq x \leq b$. 
    \item Zone~\cite{Mine:Zone}: A relational domain supporting constraints of the form $x - y \leq c$ (the difference constraints) and single-variable bounds. 
    \item Octagon~\cite{mine2006octagon}: A relational domain supporting constraints of the form $\pm x \pm y \leq c$, where $x$ and $y$ are variables and $c$ is a constant.  
    \item KnownBit: A non-relational domain that considers each bit of the bit-vector variable independently, constraining whether it must be $0$ or $1$.
\end{itemize}

Using symbolic abstraction~\cite{li2014symbolic,jiang2017block,yao2021program,reps2004symbolic}, we derive optimal domain transformers by formulating and solving corresponding Optimization Modulo Theories (OMT) queries. For each template expression $e_i$, we compute $\max \ e_i \ \text{s.t.} \ \varphi$, obtaining a constraint $e_i \leq c_i$. 

\smallskip
\noindent \textbf{Environment}.
All experiments were conducted on a 64-bit server with Intel Xeon Gold 5218R CPUs @ 2.10 GHz and 48GB of RAM. 

\subsection{Precision of Best Abstraction (RQ1)}
\label{subsec:rq1}

To evaluate precision, we measure two complementary metrics for each formula $\varphi$ and abstract domain $A$:
\begin{itemize}
    \item \emph{False positive rate} (FPR): The proportion of spurious models among all models captured by the abstraction.
    \item \emph{Precision}: Defined as $1-\text{FPR}$, indicating how closely the abstract semantics approximates the concrete semantics.
\end{itemize}

\begin{figure}[t]
    \centering
    \includegraphics[width=0.8\linewidth]{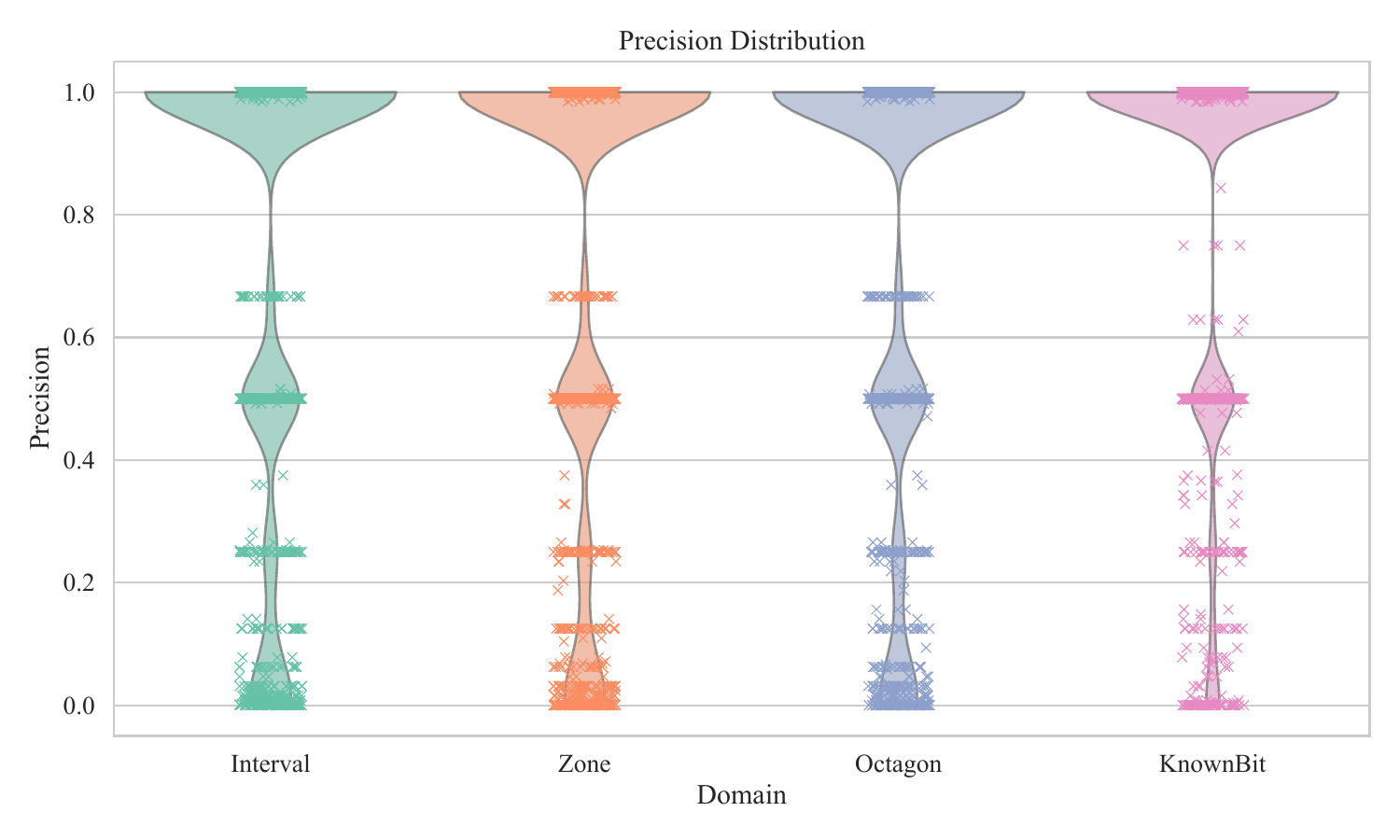}
    \caption{Precision comparison across domains: Distribution of precision scores (higher is better) for Interval, Zone, Octagon, and KnownBit domains on our benchmark set. Each point represents a formula. Note the unexpectedly small gap between the precision of Interval, Zone, and Octagon.}
    \label{fig:precision}
\end{figure}

Figure~\ref{fig:precision} illustrates the distribution of precision values for the domains. Each data point represents the precision of a domain on a single formula. The results show three main trends:

\begin{itemize}
    \item The Interval domain achieves an average precision of 76.1\%. 
     Despite its simplicity, it performs on par with more expressive numerical domains.
    \item The Zone domain reaches an average precision of 76.6\%. This domain captures relational information through various constraints, which one might expect to yield a significant improvement in precision. However, we observe that the gain over the interval is marginal—only 0.5 percentage points.
    
    \item The Octagon domain attains an average precision of 77.2\%. Despite its greater expressiveness, it does not significantly outperform simpler domains.

    \item The KnownBit domain achieves an average precision of 85.7\%, substantially higher than any word-level domain. 
\end{itemize}

\mybox{\finding\ The Interval domain achieves surprisingly high precision (76.1\%) relative to the more expressive relational domains (Zone: 76.6\%, Octagon: 77.2\%). 
}

In static analysis literature, relational domains are often favored for their ability to capture inter-variable constraints, which are presumed to improve analysis accuracy. In particular, we utilize symbolic abstraction to determine the optimal abstractions for these domains.
However, our results suggest these relational constraints may not capture significantly more useful information than simple variable bounds.

\mybox{\finding\  The KnownBit domain consistently outperforms word-level domains, achieving a mean precision of 85.7\%. This suggests that per-bit reasoning captures structural properties of formulas that numerical abstractions miss.}

KnownBit abstraction treats each bit of a variable independently, enabling it to represent constraints like ``the third bit must be 1''—a type of property that word-level numeric domains are hard to express. This capability is particularly valuable in bit-vector reasoning, where such properties often have semantic significance (e.g., when masking or shifting bits).

Another potential factor in KnownBit's superior performance is that program analysis tools often generate inherently bit-structural formulas. For instance, path constraints in KINT often originate from low-level operations, such as flags and bitfields. In such contexts, word-level numerical relations may be too coarse, while bit-level constraints align more with the actual semantics.

\subsection{Comparing Different Abstract Domains (RQ2)}
\label{subsec:rq2}

This section presents a comparative analysis of pairs of domains to identify which domains perform better overall and which provide complementary or redundant information. 

\begin{figure}[t]
    \centering
    \begin{subfigure}[b]{0.48\textwidth}
        \centering
        \includegraphics[width=\textwidth]{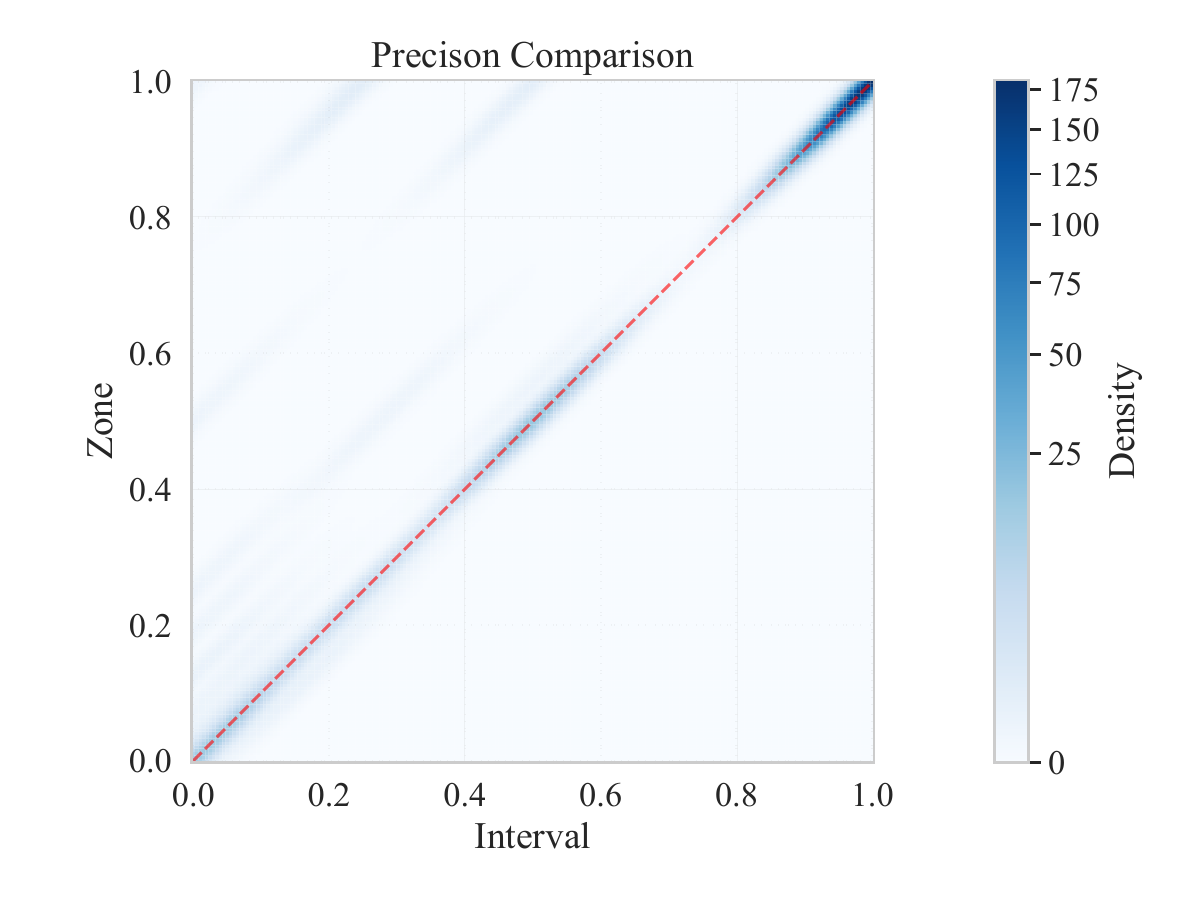}
        \caption{Interval vs Zone}
    \end{subfigure}
    \begin{subfigure}[b]{0.48\textwidth}
        \centering
        \includegraphics[width=\textwidth]{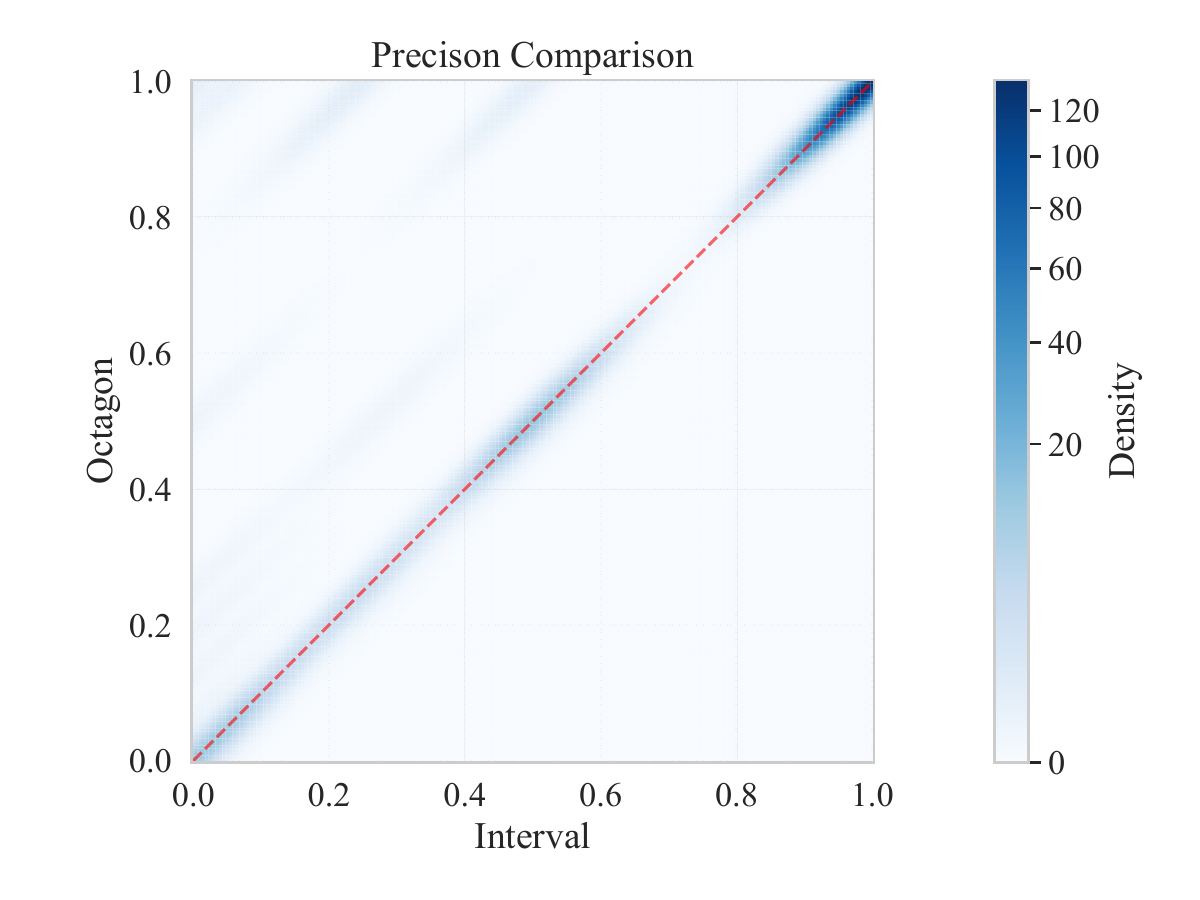}
        \caption{Interval vs Octagon}
    \end{subfigure}
    \begin{subfigure}[b]{0.48\textwidth}
        \centering
        \includegraphics[width=\textwidth]{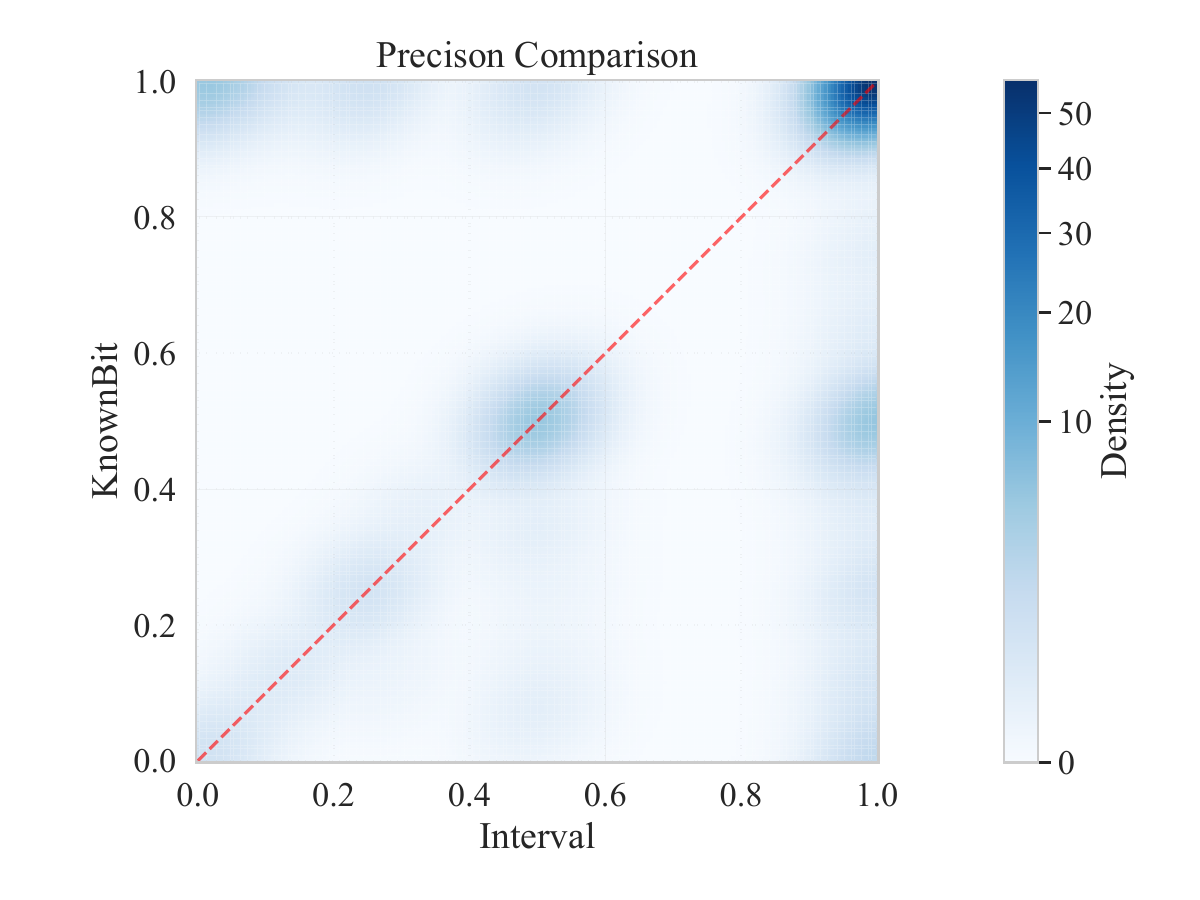}
        \caption{Interval vs KnownBit}
    \end{subfigure}
    \caption{Distribution of benchmark formulas for precision comparison across different domains. The deeper the color, the more formula falls into the corresponding region. The formulas on the $x=y$ line apply when the two domains have the same precision.}
    \label{fig:precision_comparison}
\end{figure}

To begin, we generate scatter plots where each point represents a benchmark formula. The coordinates on the plot correspond to the precision values of two domains for that formula. Figure~\ref{fig:precision_comparison} displays three such comparisons: (a) Interval vs. Zone, (b) Interval vs. Octagon, and (c) Interval vs. KnownBit. The $x = y$ diagonal line denotes equal precision for both domains on a given formula. From these plots, several trends emerge: 
\begin{itemize}
    \item Most formulas lie close to the diagonal in the Figures~\ref{fig:precision_comparison}(a)-(b), indicating that these domains tend to produce very similar levels of precision; 
    \item Few outliers suggest modest gains in precision for Zone and Octagon, but these are rare;
    \item  Many formulas appear above the diagonal in Figure~\ref{fig:precision_comparison}(c), showing that the KnownBit domain often achieves noticeably higher precision.
\end{itemize}

To quantify these observations, we introduce the notion of domain-specific false positives, i.e., models that are spurious in one domain but not in the other. This helps us identify which domain introduces more abstraction noise.
We then compute the domain-specific false-positive rate as the proportion of these exclusive false positives among all false positives in that domain.
\begin{itemize}
    \item The average Interval-specific false positive rate is 1.1\% higher than Zone, 1.3\% higher than Octagon, and 12.9\% higher than KnownBit, indicating that the Interval domain does not have many more extra false positives than Zone or Octagon, but considerably more than KnownBit.
    \item The KnownBit domain has an average specific false positive rate of 3.5\%, suggesting that KnownBit also produces some extra false positives not found in the Interval domain.
     For some formulas, the Interval-specific false positive rate is high while the KnownBit-specific rate is almost zero. Conversely, for other formulas, the KnownBit-specific false-positive rate is high, whereas the Interval-specific rate is relatively low.
\end{itemize}

\begin{figure}[t]
    \centering
    \includegraphics[width=0.75\linewidth]{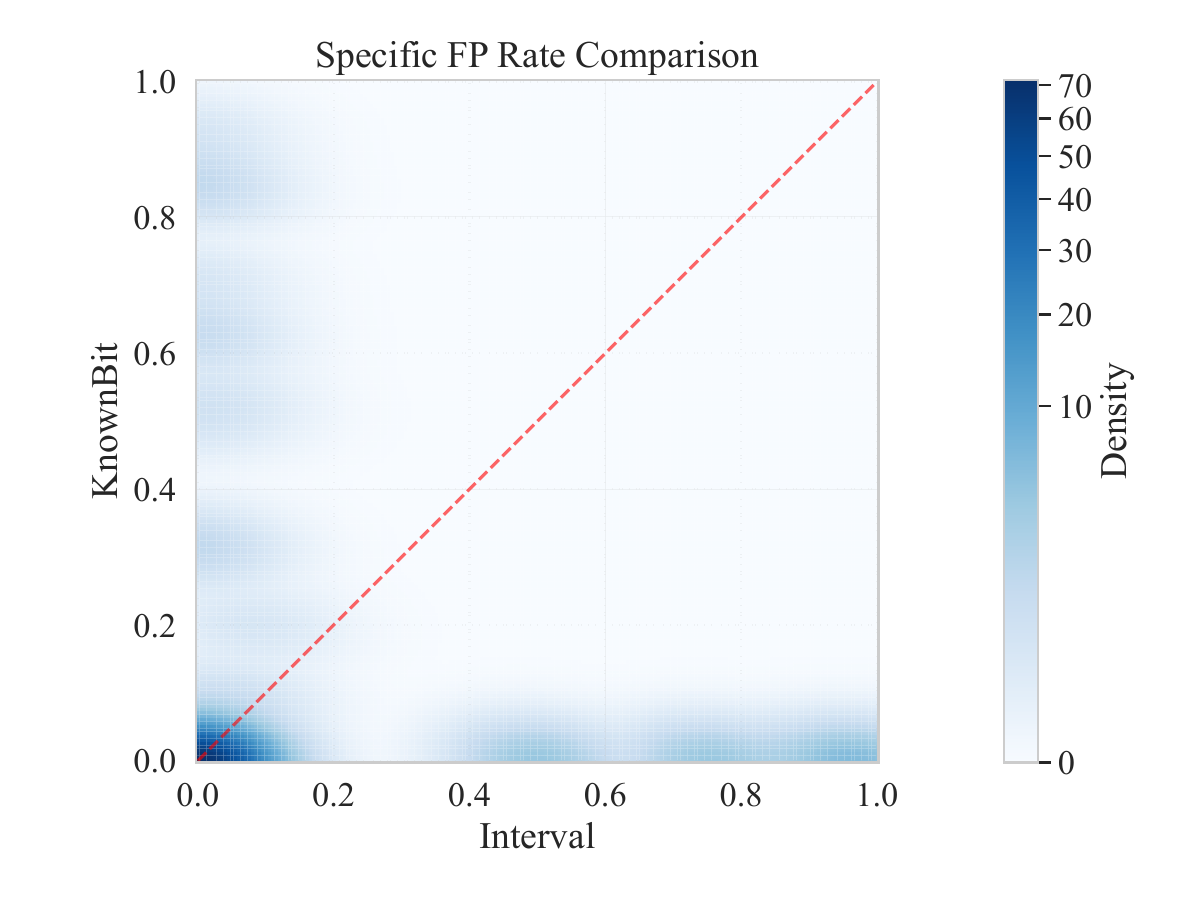}
    \caption{Specific false positive rate comparison between Interval and KnownBit: Distribution of benchmark formulas on precision comparison between the Interval and KnownBit domains. The deeper the color, the more formula falls into the corresponding region. The $x$-axis and the $y$-axis represent the domain-specific false-positive rates of Interval and KnownBit, respectively. 
    }
    \label{fig:false_positive_rate}
\end{figure}

\mybox{\finding\  The Zone and Octagon domains provide only marginal gains over Interval, with very low domain-specific false positive rates (1.1–1.3\%).}

Despite the theoretical advantages of relational domains, their practical benefits are limited in this context. Most of the abstraction precision in the Zone and Octagon domains overlaps with the Intervals. Their additional expressiveness, such as the ability to capture the difference constraints, does not translate into substantial reductions in false positives.

\mybox{\finding\  The KnownBit domain captures a substantial set of false positives not identified by Interval (12.9\%), while also introducing a few unique false positives (3.5\%). }

This finding has strong implications for the abstraction strategy. Rather than treating Interval and KnownBit as mutually exclusive options, combining them could yield a more precise composite abstraction.
It is also noteworthy that although KnownBit has some domain-specific false positives (3.5\%), the trade-off is a significant reduction in false positives compared to Interval, making KnownBit a strong candidate for improving precision without incurring high costs (as shown later in RQ4).

\subsection{Constraint Redundancy within the Octagon Domain (RQ3)}
\label{subsec:rq3}
The Octagon abstract domain is known for its ability to express linear constraints involving sums and differences of variables. While it is more expressive than simpler domains, such as Interval or Zone, this increased expressiveness comes at a computational cost. Thus, a natural question arises: do all categories of constraints within the Octagon domain contribute meaningfully to its precision, or are some types effectively redundant in practice?

To investigate this, we decomposed the Octagon domain into three constraint categories based on their structural form:

\begin{itemize}
    \item \emph{Interval}: Constraints of the form $\pm x \leq c$;
    \item \emph{Plus}: Constraints of the form $x + y \leq c$ or $-x - y \leq c$;
    \item \emph{Minus}: Constraints of the form $x - y \leq c$ or $-x + y \leq c$.
\end{itemize}

\begin{figure}[t]
    \centering
    \includegraphics[width=0.9\linewidth]{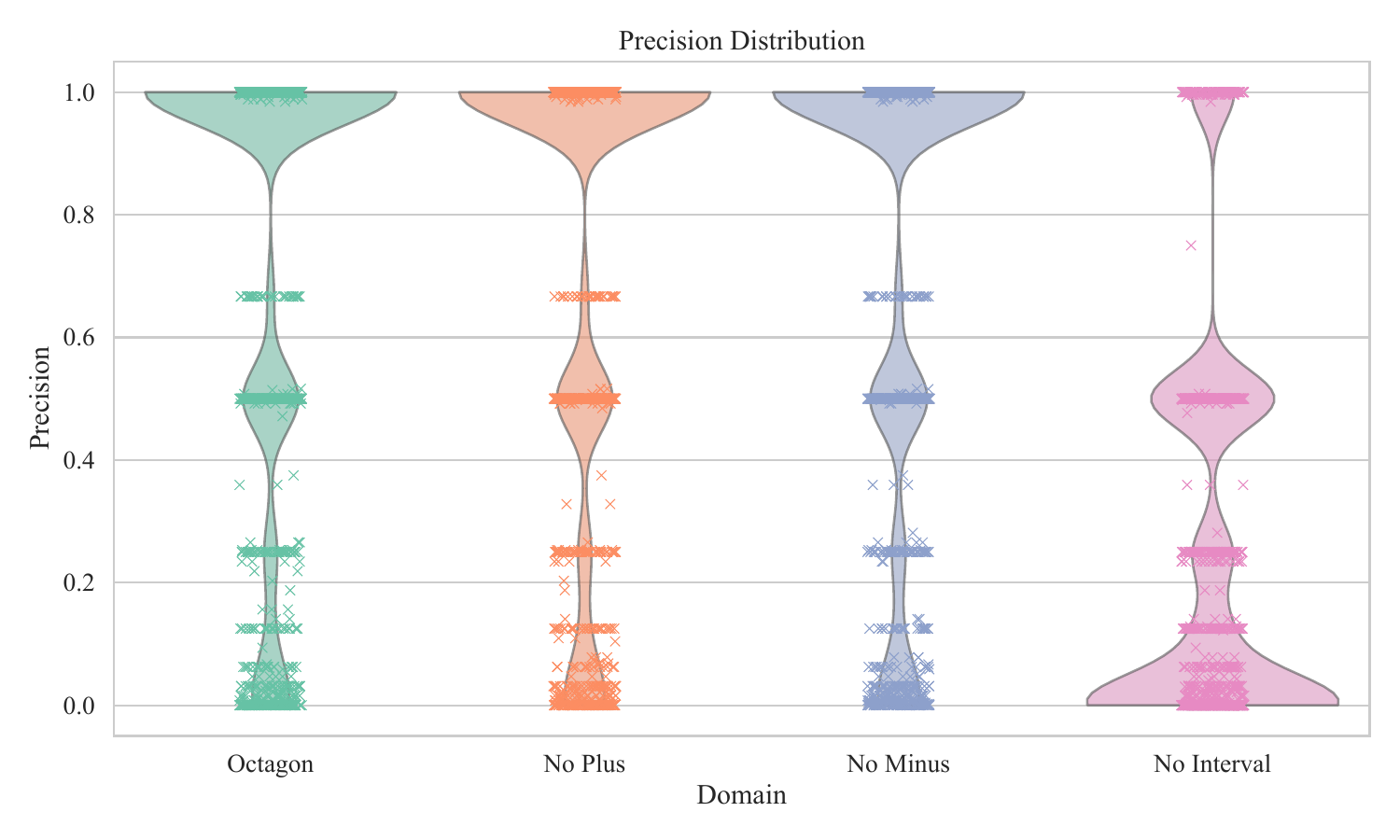}
    \caption{Impact of constraint categories in Octagon: Precision when removing different constraint types compared to the full Octagon domain. ``No Interval'' removes single-variable bounds, ``No Minus'' removes difference constraints of form $x - y \leq c$, and ``No Plus'' removes sum constraints of form $x + y \leq c$. 
    }
    \label{fig:octagon_constraints}
\end{figure}

To measure the contribution of each category, we conducted an ablation study by selectively removing a single constraint from the full Octagon abstraction and measuring the resulting change in precision. 
Figure~\ref{fig:octagon_constraints} presents the results of this analysis. The full Octagon domain achieves a mean precision of 77.2\%. When each category is removed independently, we observe the following:
\begin{itemize}
    \item Removing Interval constraints causes the most dramatic precision drop, from 77.2\% to 24.0\% (a 53.2 percentage point decrease);
    
    \item Removing Minus constraints reduces precision to 76.3\% (a 0.9 percentage point decrease);
    
    \item Removing Plus constraints has a minimal impact on precision, resulting in 76.6\% (nearly the same as the full Octagon domain).
\end{itemize}

\mybox{\finding\  Plus constraints in Octagon domain contribute negligibly to precision improvement. In contrast, Interval constraints account for most of the domain precision gains, while Minus constraints provide limited additional value.}

This finding has practical implications for the design and implementation of abstract domains. It suggests that much of the computational effort spent maintaining the plus constraints in Octagon abstractions may not be justified by a corresponding gain in accuracy. In many cases, simply combining Interval and Minus constraints—effectively forming a streamlined version akin to the Zone domain—may suffice.

\subsection{Performance Analysis of MCAI (RQ4)}
\label{subsec:rq5}
In this subsection, we evaluate the computational cost of MCAI across different abstract domains.
We measured two key performance metrics:
\begin{itemize}
    \item \emph{Abstraction Time}: The time required to compute the best abstraction via symbolic abstraction.
    \item \emph{Counting Time}: The time required to perform model counting.
\end{itemize}

\begin{table}[t]
    \centering
    \caption{Performance across different domains (times in seconds)}
    \label{tab:performance}
    \begin{tabular}{llcccc}
    \hline
    \multicolumn{2}{c}{\textbf{Metric}} &\textbf{Interval} & \textbf{Zone} & \textbf{Octagon} & \textbf{KnownBit} \\ \hline
    \multirow{4}{*}{Abstraction Time} & Min & 0.01 & 0.00 & 0.00 & 0.21 \\
    & Max & 0.25 & 132.61 & 610.06 & 5.18 \\
    & Median & 0.04 & 0.16 & 0.25 & 0.99 \\
    & Average & 0.05 & 3.84 & 18.07 & 1.23 \\ \hline
    \multirow{4}{*}{Counting Time} & Min & 0.31 & 0.32 & 0.29 & 0.32 \\
    & Max & 72.81 & 162.19 & 166.56 & 72.83 \\
    & Median & 0.44 & 0.46 & 0.60 & 0.46 \\
    & Average & 5.53 & 32.78 & 49.91 & 5.76 \\ \hline
    \multirow{4}{*}{Total Time} & Min & 0.32 & 0.33 & 0.35 & 0.59 \\
    & Max & 72.94 & 267.17 & 745.44 & 76.16 \\
    & Median & 0.48 & 0.76 & 4.65 & 1.49 \\
    & Average & 5.58 & 36.62 & 67.98 & 6.99 \\ \hline
    \end{tabular}
\end{table}

Table~\ref{tab:performance} summarizes the performance characteristics across different domains.

The analysis time increases significantly with domain expressiveness. The Interval domain, being the simplest, is the most efficient, with a total median analysis time of 5.58 seconds per formula. In contrast, the Octagon domain, which includes both difference and additive constraints between variables, is the slowest, taking 67.98 seconds on average, over 12$\times$ slower than Interval.

The Zone domain represents a middle ground. It is more expressive than Interval but less complex than Octagon, and it achieves a median total time of 36.62 seconds—about 6.5$\times$ slower than Interval.
Interestingly, the KnownBit domain—despite its high precision—achieves a total analysis time of just 6.99 seconds, only slightly above that of Interval. This makes it more accurate (as seen in RQ1 and RQ2) and computationally efficient.

Table~\ref{tab:performance} also reveals an interesting pattern: while the Zone and Octagon domains have median analysis times only modestly higher than the Interval domain (0.76s and 4.65s vs 0.48s), their maximum times are dramatically larger (267s and 745s vs 73s). This suggests that relational domains do not always incur prohibitive computational costs—in many cases, their overhead is more acceptable than our expectations given the potential precision gains they offer. The extreme maximum times likely correspond to particularly complex formulas where the additional expressiveness leads to state explosion.

\mybox{\finding\  Octagon incurs a 12$\times$ overhead compared to Interval, while Zone is 6.5$\times$ slower. KnownBit achieves higher precision with only a 1.25$\times$ increase in runtime.}

%% file: 6.discuss.tex
\section{Discussions}
\label{sec:discuss}

Several directions remain for applying and extending MCAI, both to analyze abstract interpreters and to improve their design.

\smallskip
\noindent\textbf{Evaluating Non-optimal Transformers.}
We focus on the best abstractions, which represent the ideal precision achievable in a domain. In practice, transfer functions are often approximate due to performance trade-offs, heuristics, or solver limitations. MCAI can quantify the precision gap between an implemented transformer and the domain's best abstraction for the same concrete semantics. Given a concrete formula $\varphi$ and an analyzer-produced abstract element $a$ (or its symbolic concretization $\varphi_a$), MCAI can compute (i) the false positive rate of $\varphi_a$ with respect to $\varphi$, and (ii) the additional imprecision relative to $\alpha_A(\varphi)$.These metrics support fine-grained diagnosis of precision bottlenecks. For instance, MCAI enables differential testing across analyzer versions or configurations; it can flag regressions and help localize precision loss.

\smallskip
\noindent\textbf{Evaluating Redundancies in Fixpoint Iteration.}
Prior work~\cite{he2020learning} has shown that fixpoint iteration often introduces constraints that are not necessary for
proving the final invariant, especially when joins and widenings accumulate redundant information.
MCAI can be adapted to quantify such redundancies by measuring the marginal contribution of
constraints to the overall reduction in false positives.
For example, for a sequence of iterates $a_0 \sqsubseteq a_1 \sqsubseteq \cdots \sqsubseteq a_k$ (or the
constraints constituting a single iterate), MCAI can measure how much each step reduces the spurious-model set relative to the concrete semantics.
This yields an empirical basis for simplifying invariants, compressing intermediate states, or
designing more effective join/widening strategies that preserve ``useful'' constraints while discarding
those with negligible semantic impact.

\smallskip
\noindent\textbf{Beyond Numerical Bit-Vectors.} MCAI requires a symbolic concretization and a model-counting mechanism over the relevant state space. This generality allows application to other finite or bounded domains, such as typestate automata, bit-precise pointer metadata, or bounded heaps. For infinite-state domains, MCAI can be applied via principled bounding techniques (e.g., bit-width restrictions). Extending MCAI to these settings offers a systematic approach to precision measurement across a broader class of analyses.

%% file: 7.related.tex
\section{Related Work}
\label{sec:related}

\noindent \textbf{Evaluating Abstract Domains}.
Several researchers have addressed the challenge of quantifying precision loss in abstract interpretation. 
\citet{logozzo2009towards} evaluated numerical domains by comparing the minimal set of linear constraints that encode their abstract values; smaller sets indicate greater precision. 
\citet{di2008relational} compared various relational domains using specific benchmark programs, measuring their ability to verify array-bound checks and related properties. 
Similarly, \citet{rountev2004evaluating} quantitatively compared static analyses for specific client applications. 
\citet{sotin2010quantifying} proposed a framework that measures interval domain precision by comparing the volume of the abstract element with the volume of the concrete set it represents. 
However, none of these approaches leverage model counting to directly measure the semantic gap between abstract elements and concrete semantics as MCAI does.

\smallskip
\noindent \textbf{Completeness in Abstract Interpretation}.
The concept of completeness in abstract interpretation was first explored by \citet{cousot1979systematic}, who distinguished between sound approximations and complete ones.
\citet{giacobazzi1997completeness} provided a constructive characterization of complete abstract interpretations and introduced property-specific notions of completeness.
\citet{giacobazzi2000making} further developed techniques to systematically refine abstract domains to achieve completeness with respect to specific properties. More recently, \citet{bruni2021logic} explored completeness for specific logical fragments, showing how to design domains that precisely capture certain classes of properties.
Our work contributes to this line of research by quantifying incompleteness via model counting. 
Rather than a binary notion of completeness, MCAI offers a spectrum that quantifies the degree of incompleteness.

\smallskip
\noindent \textbf{Model Counting in Program Analysis}.
Model counting solvers~\cite{sharpsat,approxmc,trinh2017model,eiers2019subformula,fredrikson2014satisfiability} have emerged as valuable tools in program analysis~\cite{fredrikson2014satisfiability,eiers2019subformula,luu2014model}, enabling quantitative reasoning about program behaviors. These solvers have been applied to a range of domains, including probabilistic symbolic execution~\cite{borges2015iterative,aydin2015automata,aydin2018parameterized}, quantitative information flow~\cite{backes2009automatic,bang2016string}, and automated attack synthesis~\cite{bang2018online}. 
For example, we can quantify the probability of satisfying a path condition or measure information leakage under a given observation model. 
A recent survey by \citet{tevfik2020quantifying} provides a comprehensive overview of these applications. MCAI extends this line of work by introducing a quantitative notion of semantic precision for abstract domains. It evaluates and compares abstractions by measuring their divergence from the concrete semantics.

%% file: 8.conclu.tex
\section{Conclusion}
\label{sec:conclu}
Rigorous evaluation is essential for advancing static analysis.
We have presented MCAI, a quantitative approach for evaluating the precision of abstract domains with respect to the concrete semantics. We have applied MCAI to four abstract domains, demonstrating its utility for systematically comparing domain precision and providing insights into designing new abstract domains.